\documentclass[]{aa}       
\usepackage{graphicx}
\usepackage{amssymb}
\usepackage{longtable}
\usepackage{supertabular}
\usepackage{natbib} 
\usepackage{ulem}

\begin{document}

\title{Simulations of stellar/pulsar-wind interaction along one full orbit}

\author{
V. Bosch-Ramon \inst{1} \and
M.V. Barkov\inst{2,3} \and
D. Khangulyan\inst{4} \and
M. Perucho\inst{5}
}

\authorrunning{Bosch-Ramon et al.}

\titlerunning{Two-dimensional simulations of interacting stellar/pulsar winds}

\institute{
Dublin Institute for Advanced Studies, 31 Fitzwilliam Place, Dublin
2, Ireland; valenti@cp.dias.ie
\and
Max Planck Institut f\"ur Kernphysik, Saupfercheckweg 1, Heidelberg 69117,
Germany
\and
Space Research Institute, 84/32 Profsoyuznaya Street, Moscow, Russia
\and
Institute of Space and Astronautical Science/JAXA, 3-1-1 Yoshinodai, Chuo-ku, Sagamihara, Kanagawa 252-5210, Japan; khangul@astro.isas.jaxa.jp
\and
Dept. d'Astronomia i Astrof\'{\i}sica, Universitat de Val\`encia, C/ Dr. Moliner 50, 46100, Burjassot (Val\`encia), Spain; Manel.Perucho@uv.es 
}

\offprints{V. Bosch-Ramon, \email{valenti@cp.dias.es}}

\date{Received <date> / Accepted <date>}

\abstract
{The winds from a non-accreting pulsar and a massive star in a binary system collide forming a bow-shaped shock structure.
The Coriolis force induced by orbital motion deflects the shocked flows, strongly affecting their dynamics.}     
{We study the evolution of the shocked stellar and pulsar winds on scales in which the orbital motion is important. Potential
sites of non-thermal activity are investigated.}    
{Relativistic hydrodynamical simulations in two dimensions, 
performed with the code {\it PLUTO}{} and using the adaptive mesh
refinement technique, are used to model interacting stellar and pulsar winds on scales $\sim 80$ times the distance between
the stars. The hydrodynamical results suggest the suitable locations of sites for particle acceleration and non-thermal
emission.}    
{In addition to the shock formed towards the star, the shocked and unshocked components of the pulsar wind flowing away from
the star terminate by means of additional strong shocks produced by the orbital motion. 
Strong instabilities lead
to the development of turbulence and an effective two-wind mixing in both the leading and trailing sides of the interaction
structure, which starts to merge with itself after one orbit. The adopted moderate pulsar-wind 
Lorentz factor already provides a good qualitative description of the phenomena involved in high-mass 
binaries with pulsars, and can capture important physical effects that would not appear in non-relativistic treatments.}  
{Simulations show that shocks, instabilities, and mass-loading yield efficient mass, momentum, and energy exchanges between the
pulsar and the stellar winds. This renders a rapid increase in the entropy of the shocked structure, which will likely be 
disrupted 
on scales beyond the simulated ones. Several sites of particle acceleration and low- and high-energy emission 
can be identified. Doppler boosting will have significant and complex effects on radiation.} 

\maketitle

\section{Introduction} \label{intro}

The binary system PSR~B1259$-$63/LS2883, consisting of a late O star \citep{neg11} and a 47~ms pulsar \citep{joh92}, is a
powerful GeV and TeV emitter \citep[e.g.][]{aha05,abd11,tam11}. The physical processes behind the gamma-ray emission detected
in PSR~B1259$-$63/LS2883 are believed to take place in the region where the stellar and pulsar winds collide, which would
also produce X-rays and, once the interacting flows have left the system, radio emission
\citep[e.g.,][]{tav97,jmm99,dub06,mol11}. The processes underlying the non-thermal radiation in PSR~B1259$-$63/LS2883 seem to
be shocks that take place in the two-wind interaction region and accelerate electrons. These electrons most likely radiate
from radio-to-X-rays through synchrotron and at GeV and TeV energies through inverse Compton emission  (IC), although some
gamma rays may come from IC in the unshocked pulsar wind \citep[e.g.,][]{tav97,bk00,kha07,kha11,kha12}. The details of the
dynamics of the interacting flows are still poorly known, which affects the interpretation of the observed variability and
spectra not only in PSR~B1259$-$63/LS2883 but also in other gamma-ray emitting binaries that may host a non-accreting pulsar
(LS~5039 and LS~I~+61~303, see discussions in, e.g., \citealt{che06,dub06,rom07,bk09,tor11,bed11,bar12}; HESS~J0632$+$057,
see, e.g., \citealt{bon11}; 1FGL~J1018.6$-$5856, see, e.g., \citealt{ack11,abr12}). 

Hydrodynamical and magnetohydrodynamical, relativistic and non-relativistic simulations of stellar/pulsar-wind interactions
have been carried out in the past few years. In \cite{rom07} and \cite{oka11}, three-dimensional (3D), non-relativistic
hydrodynamical (SPH) simulations were carried out to study, on binary scales but including orbital motion, the stellar/pulsar-wind interaction in LS~I~+61~303 and PSR~B1259$-$63/LS2883. In \cite{oka11}, the authors explicitly studied the interaction
with the equatorial disc present in the latter. \cite{bog08,bog12} performed axisymmetric, relativistic,
hydrodynamical, and magnetohydrodynamical simulations of the stellar/pulsar-wind interaction, with the symmetry axis joining
the two stars. The orbital motion was not accounted for, although the impact of the pulsar wind anisotropy was studied.
Therefore, from previous work the flow evolution seems to be well-understood on scales up to a few pulsar/star separation
distances ($a$). One can conclude that the shape of the interaction region depends on the pulsar-to-star wind-momentum rate
ratio ($\eta$), the shocked pulsar wind reaccelerates when flowing away from the shock facing the star, and the bending of the
interaction structure due to orbital motion is expected. 

An analytical study of the shocked flow evolution on scales larger than $a$ indicated that Coriolis forces related to the
orbital motion could amplify the shocked flow bending, terminating the pulsar wind flowing away from the star with a
strong shock, and enhance instabilities and mixing in the flow contact discontinuity (CD) \citep{bb11}. The basic idea is
that under typical $\eta$-values, $\sim 0.03-0.3$, the
kinetic luminosity of the pulsar wind dominates that of the star by a factor $\chi\approx
30(\eta/0.1)(v_{\rm w}/2\times 10^8\,{\rm cm~s}^{-1})^{-1}$, where $v_{\rm w}$ is the stellar wind velocity. Such a high
$\chi$-value, the interaction geometry, and the impact of the Coriolis force will eventually lead to the formation of a
partially confined, high entropy flow with a strong pressure gradient outwards. This will lead to isotropization of the flow
mass, momentum, and energy fluxes, and to reexpansion, with the consequent loss of structure. In this context, non-relativistic
two-dimensional (2D) simulations done by \cite{lam12} showed that the shocked flow may keep a spiral-like shape up to large
distances, although the adopted light-to-dense wind luminosity ratios, $\chi\le 1.25$, and the non-relativistic nature of the
simulations may prevent one from generalizing this conclusion.

To study in detail the pulsar/star wind interaction on scales $\gg a$, plus a pulsar wind velocity $v\rightarrow c$, we
performed 2D relativistic hydrodynamical simulations including orbital motion, where the flow is homogeneous and perpendicular
to the simulated plane. We adopted $\chi=30$ and $\eta=0.3$, the latter resulting in a geometry of the interaction
region similar, on spatial scales $\sim a$, to the 3D case for $\eta=0.1$. For complementarity, we also explored
the case with $\chi=60$ and $\eta=0.6$. We note that for $\eta<1$ and distances to the star $r\sim a$, the pulsar wind gets
diluted faster than the stellar wind in 3D than 2D. This means that in 3D, the shocked winds flowing away from the system
are deflected by the Coriolis force at a shorter distance than obtained in the present calculations. Moreover, instabilities
are typically less disruptive in 2D. All this shows that the flows simulated here are likely to be more stable than in 3D
simulations, which should be carried out in the near future. 

The paper is organised as follows: in Sect.~\ref{dyn} we present simulations of the orbital evolution of
the flow structure; in Sect.~\ref{w}, the impact of the value of the pulsar-wind Lorentz factor is studied, and
the consequences of adopting a relativistic or a non-relativistic approach discussed; finally, in
Sect.~\ref{disc} we summarize our work and present a discussion of the implications of our results for the
non-thermal processes in high-mass binaries hosting a young pulsar.

\section{Numerical simulations}\label{dyn}

\subsection{Numerical set-up}

The simulations were implemented in 2D with the {\it PLUTO} code\footnote{Link http://plutocode.ph.unito.it/index.html}
\citep{mbm07}, the piece-parabolic method (PPM) \citep{cw84}, an HLLC Riemann Solver \citep{mig05}, and using
AMR\footnote{Adaptive mesh refinement technique} through the {\it Chombo} code\footnote{Link
https://commons.lbl.gov/display/chombo/} \citep{chombo2009}. {\it PLUTO} is a modular Godunov-type code entirely written in C
intended mainly for astrophysical applications and high Mach number flows in multiple spatial dimensions.  The simulations
were run through the MPI (message passing interface) library in the cluster of Moscow State University {\it Chebyshev}.

The simulated flows were approximated as an ideal, relativistic adiabatic gas with no magnetic field, one particle
species, and a polytropic index of 4/3. The adopted resolution was $96 \times 96$ cells and 6 levels of AMR, which gives an
effective resolution of $6144 \times 6144$ cells. The size of the domain was $x \in [0,80\,a]$ and  $y \in [0,80\,a]$. 

The stellar and pulsar winds were assumed to be isotropic and moving at a constant speed (see below). We thus assumed that the
winds had already been accelerated, and did not consider the several effects affecting the state of the two winds before colliding:
the ionization/heating effects of X-rays from the shocked winds on the stellar wind formation region \citep[e.g.][]{blo94};
the pressure of the radiation from either the pulsar or the shocked winds exerted on the stellar-wind \citep[e.g.][]{gay97}\footnote{This effect is expected to
be minor due to a star luminosity that is much higher than the spin-down one.}; the effects of gravity on slow and dense material
within the binary \citep[e.g.][]{blo91,oka11}, which are particularly relevant in Be and/or very close systems; and the impact of the
stellar radiation field on the pulsar-wind Lorentz factor through Compton braking \citep[e.g.][]{bk00,kha07}. We note that,
even though all these factors may play a major role in particular sources and should be accounted for in more refined treatments
of the problem, they can be considered at this stage higher order effects and have therefore been neglected.

A pulsar-wind Lorentz factor $\Gamma=2$ was adopted because of resolution limitations. This is smaller than the conventional
value $\Gamma\sim 10^4-10^6$ \citep[see][and references therein]{kha12,abk12}, but high enough to reproduce the wind velocity
contrast in the bow-like shock region, where instabilities trigger, and to capture important relativistic effects.
Similarities and differences between the $\Gamma$-value adopted here, a more realistic one, and the non-relativistic case,
are discussed in Sect.~\ref{w}.

The orbit was assumed to be circular, with a period $T=3\times 10^5$~s and radius $a=3\times 10^{12}\,{\rm cm}$, and $v_{\rm
w}=0.01\,c$, which is typical for the isotropic/polar wind in massive stars. A slow equatorial wind is not considered here \citep[see
otherwise][]{oka11,bb11}. For the given $\eta$, $\Gamma$, and $v_{\rm w}$ values, the stellar and pulsar-wind momentum rates,
normalized to the stellar mass-loss rate $\dot{M}_{-7}=(\dot{M}/10^{-7}\,M_\odot\,{\rm yr}^{-1})$, are $\approx
2\times10^{27}\dot{M}_{-7}$ and $6\times10^{26}\dot{M}_{-7}$~g~cm~s$^{-2}$, respectively, and the pulsar-wind luminosity
$\approx 9\times 10^{36}\,\dot{M}_{-7}$~erg~s$^{-1}$. The numerical solution obtained can be used for other orbits with
similar values of $T/a$. For high $\dot{M}$- and small $a$-values, the shocked stellar wind may be radiative instead of
adiabatic, enhancing the growth of the flow instability at the CD \citep[see, e.g.,][for two non-relativistic
winds]{pit09}.  

At the beginning of the simulation, the unperturbed stellar and pulsar winds occupied the two grid halves, left and right,
respectively. Both winds were supersonic, of negligible temperature, and had a cylindrical symmetry around axes perpendicular to
the simulation plane and crossing the wind sources. The given values of $v_{\rm w}$ and $\Gamma$, together with $\dot{M}$ and $\eta$,
determine the wind densities. The star and the pulsar were initially located at $(39.5a,40a)$ and $(40.5a,40a)$, respectively, and
the orbital motion, occurring in the plane $XY$, was set counter-clockwise. The simulations included neither the magnetic field nor
an anisotropic pulsar wind, since their impact is expected to be small \citep{bog12}.

\subsection{Results}

\begin{figure}
\vspace{-0.3cm}
\includegraphics[width=0.48\textwidth,angle=0]{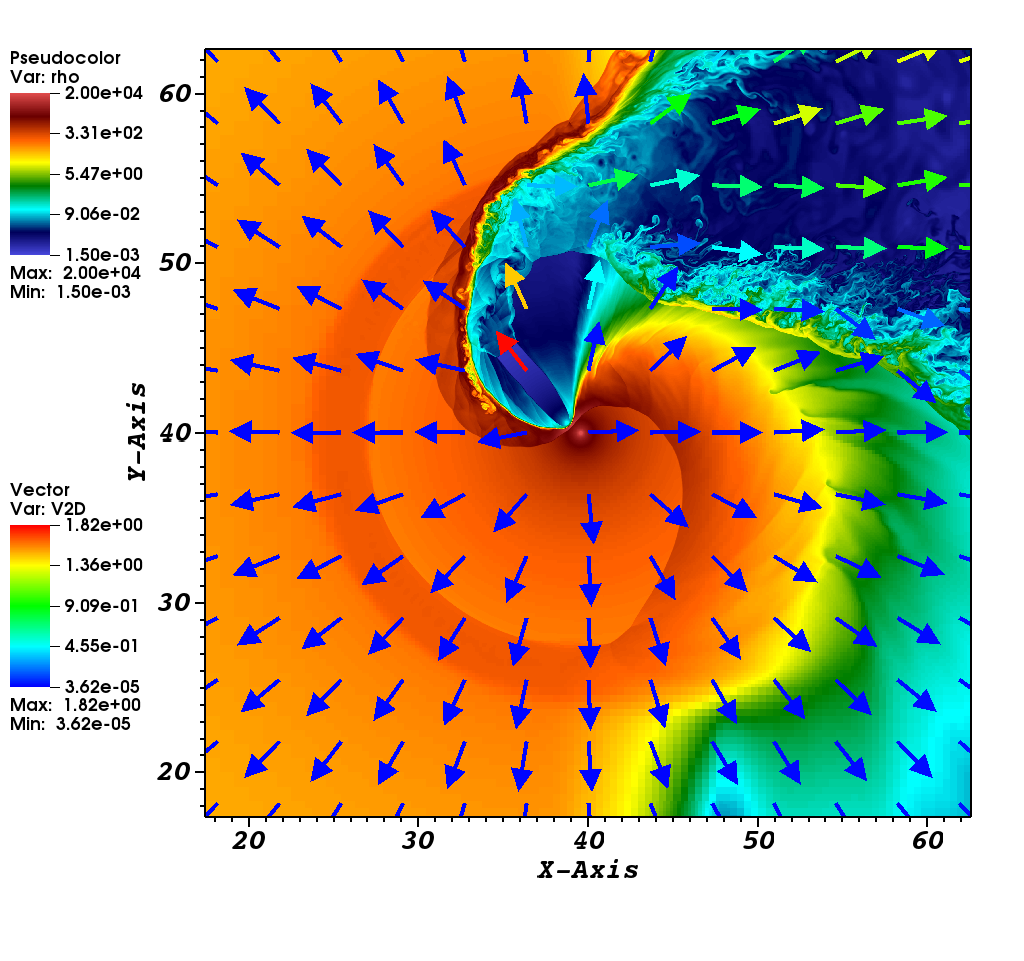}
\includegraphics[width=0.48\textwidth,angle=0]{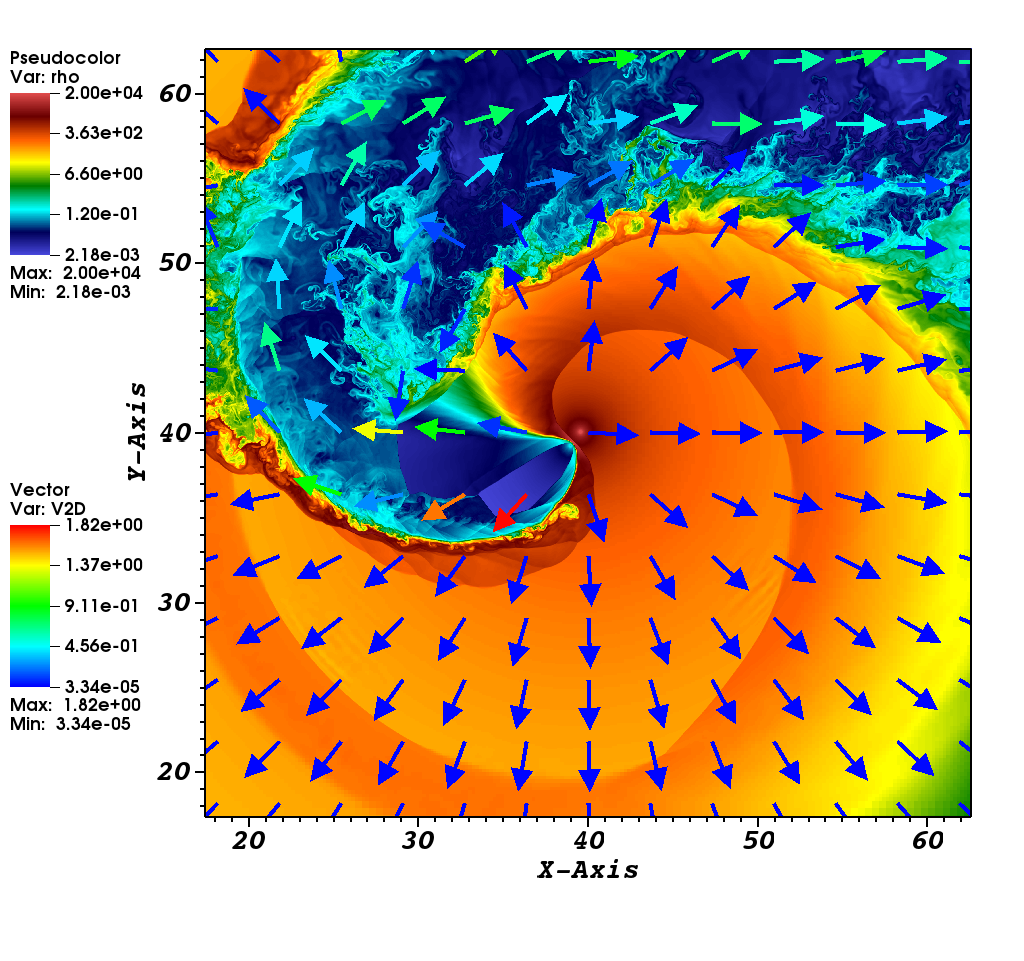}
\includegraphics[width=0.48\textwidth,angle=0]{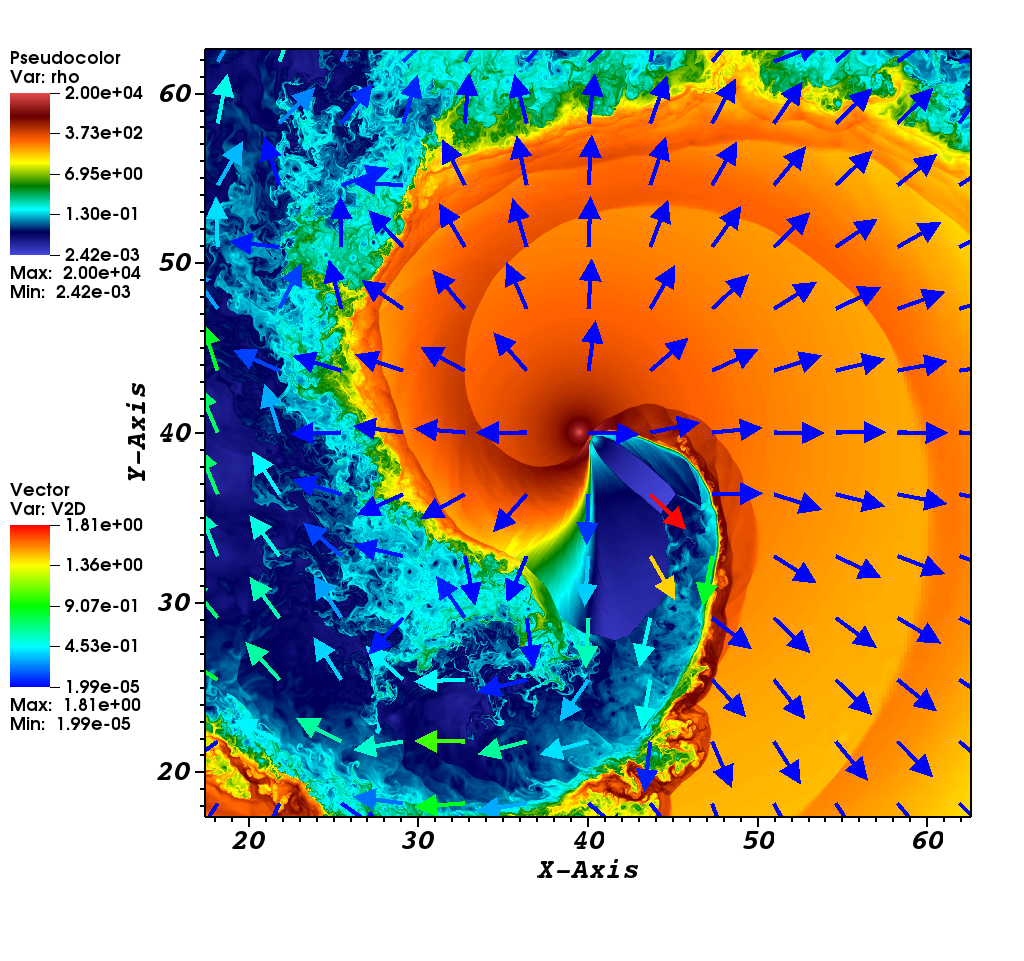}
\vspace{-1cm}
\caption{Density distribution by colour, in arbitrary units, at times
$t=1.2\times 10^5$ (top), $2.1\times 10^5$ (middle) and $2.7\times
10^5$~s (bottom), for the case with $\eta=0.3$.  The colored arrows show the four-velocity space
component for different locations, and the axes units are $a$, which
applies to all figures.}
\label{rho1}
\end{figure}

\begin{figure}
\includegraphics[width=0.5\textwidth,angle=0]{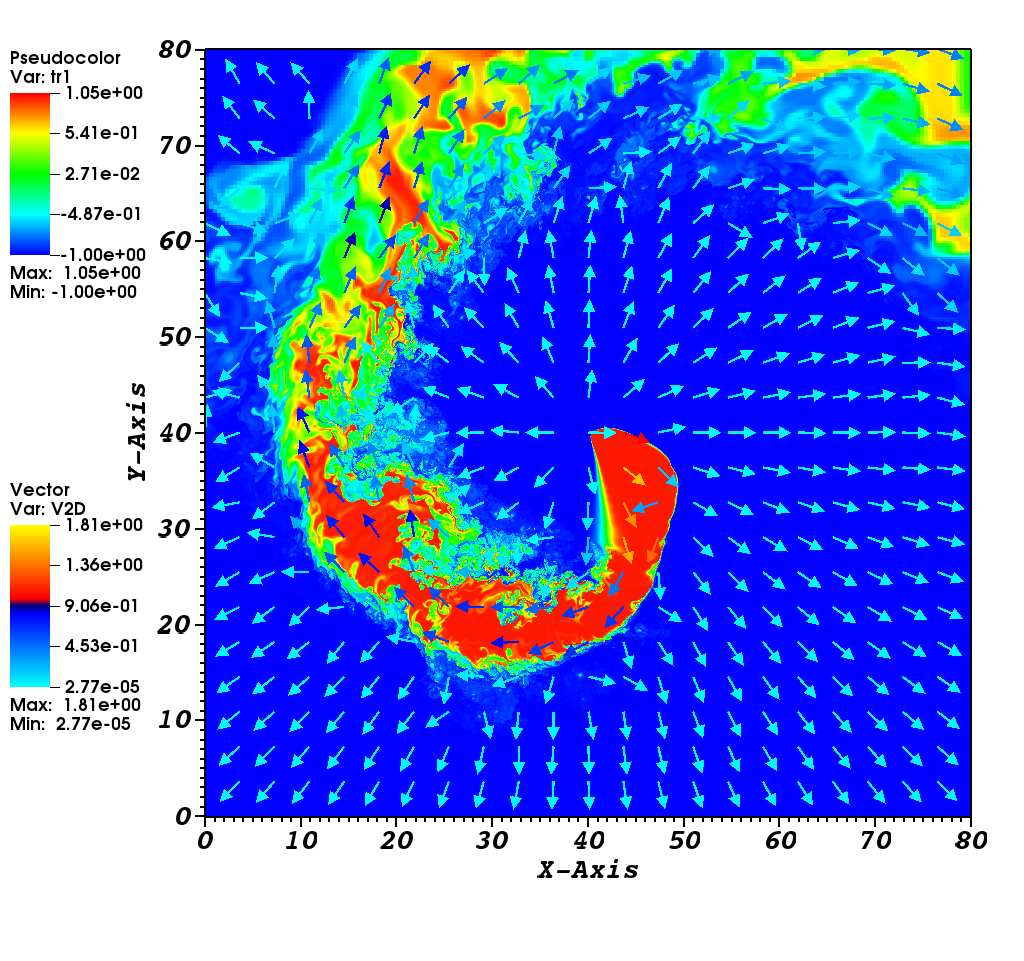}
\vspace{-1cm}
\caption{Tracer distribution by colour, where 1 is only pulsar wind and 
-1 only stellar wind, at $t=2.7\times 10^5$~s, for the case $\eta=0.3$.}
\label{rho2}
\end{figure}

\begin{figure}
\includegraphics[width=0.5\textwidth,angle=0]{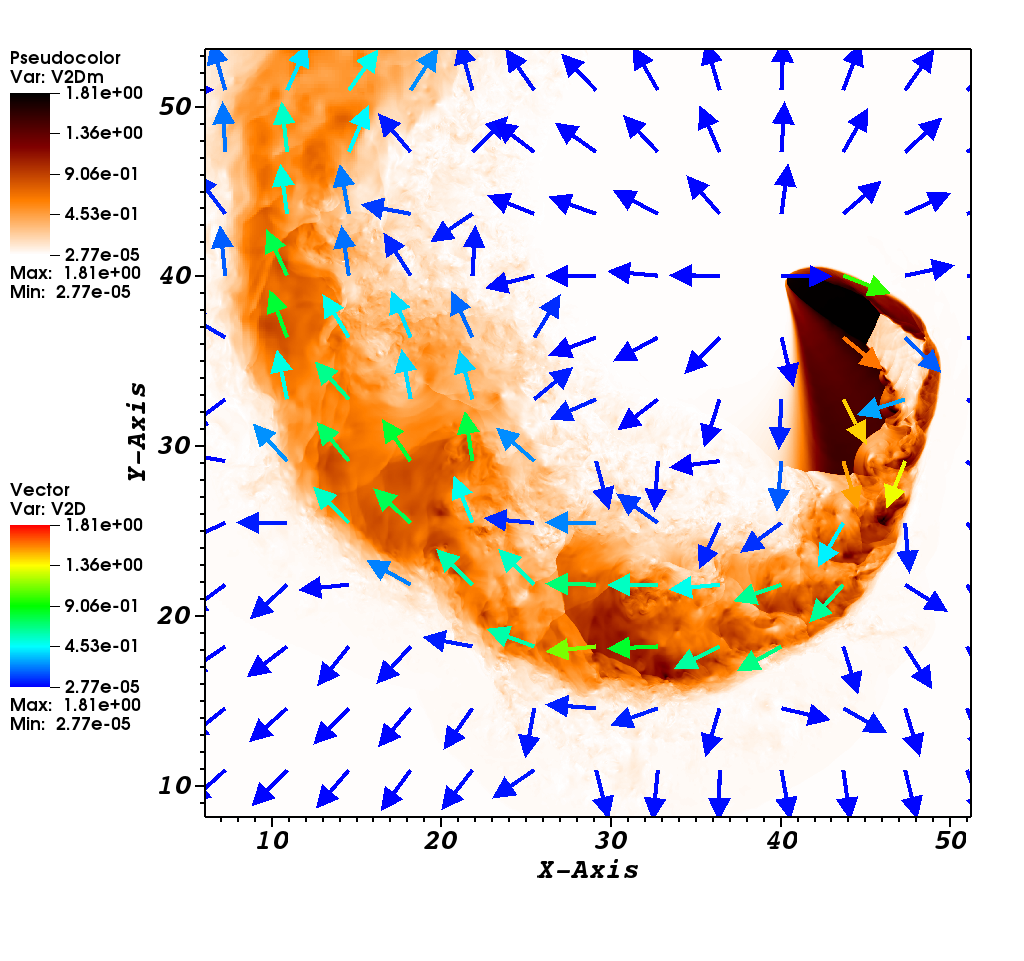}
\vspace{-1cm}
\caption{Distribution by colour of the module of the spatial component 
of the four-velocity at $t=2.7\times 10^5$~s, for the case $\eta=0.3$.}
\label{rho3}
\end{figure}

\begin{figure}
\vspace{-0.3cm}
\includegraphics[width=0.48\textwidth,angle=0]{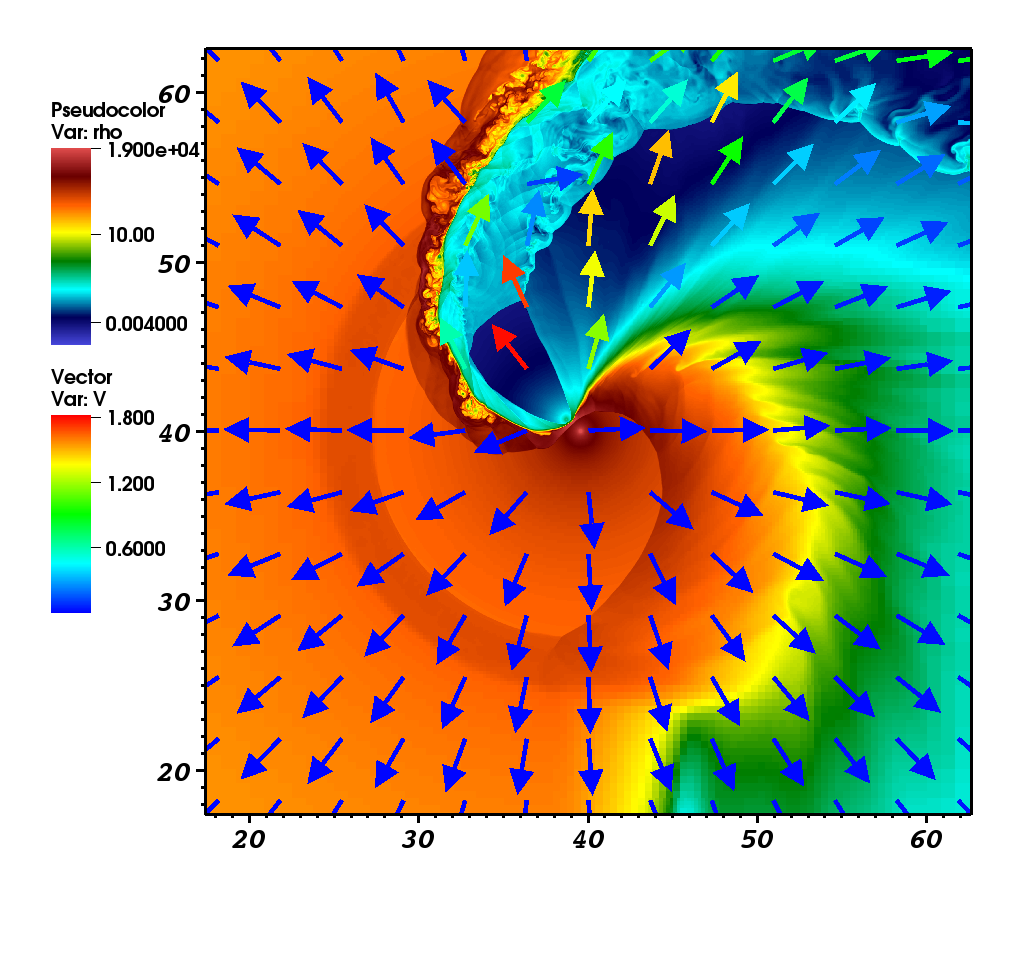}
\includegraphics[width=0.48\textwidth,angle=0]{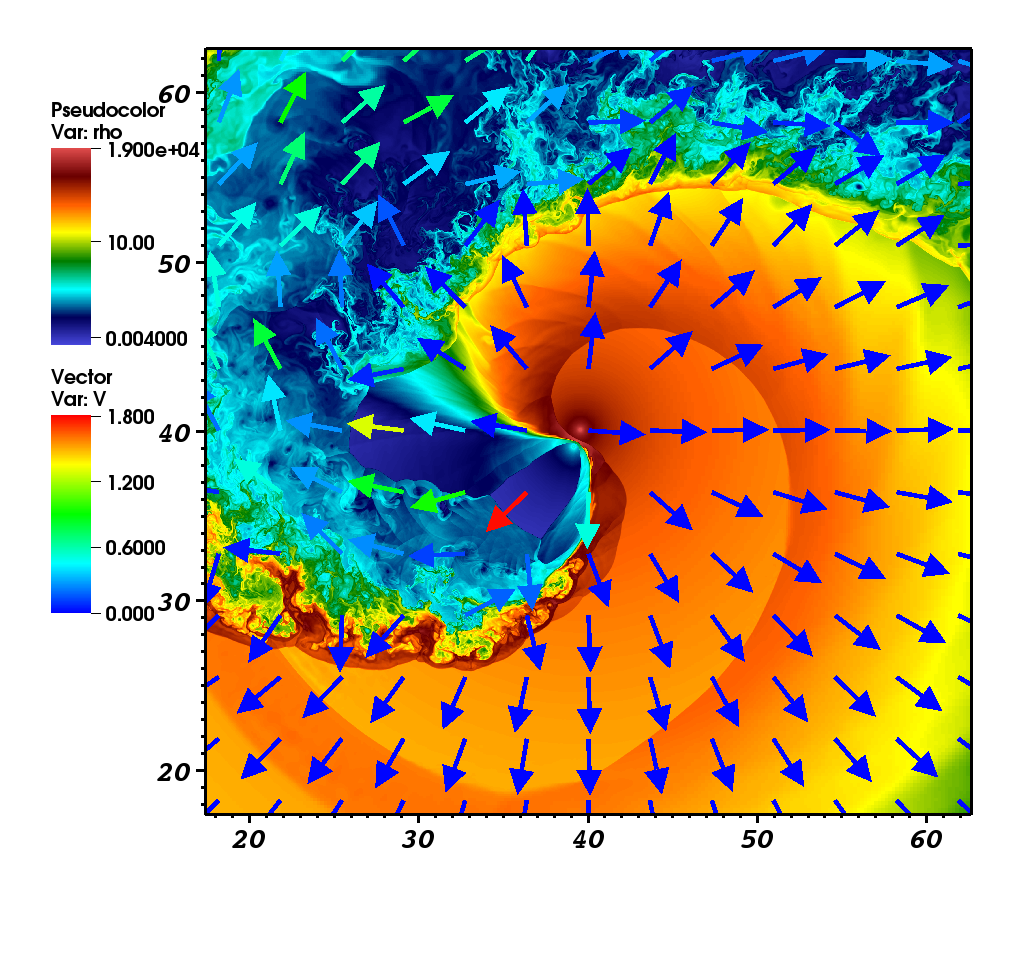}
\includegraphics[width=0.48\textwidth,angle=0]{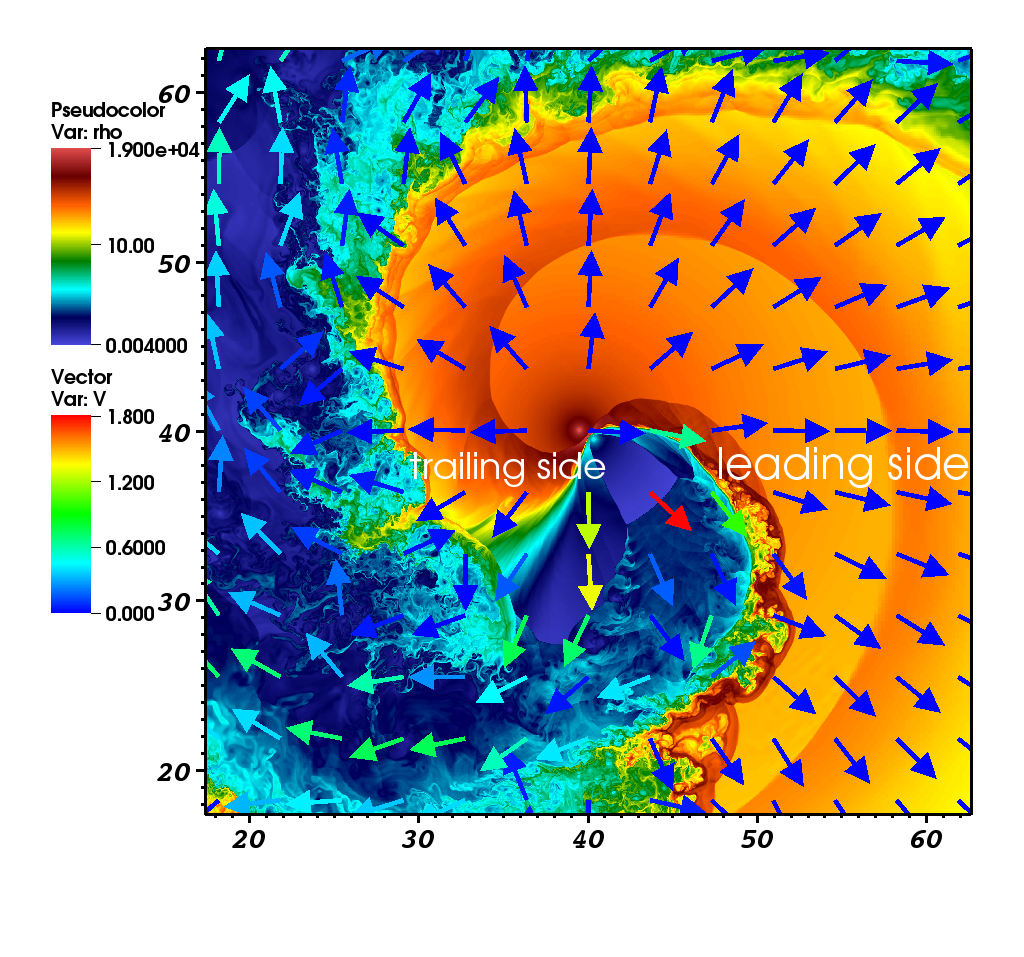}
\vspace{-1cm}
\caption{The same as in Fig.~\ref{rho1} but for the case $\eta=0.6$.
The trailing and the leading sides of
the interaction structure are indicated at the bottom panel.}
\label{rho12}
\end{figure}

\begin{figure}
\includegraphics[width=0.5\textwidth,angle=0]{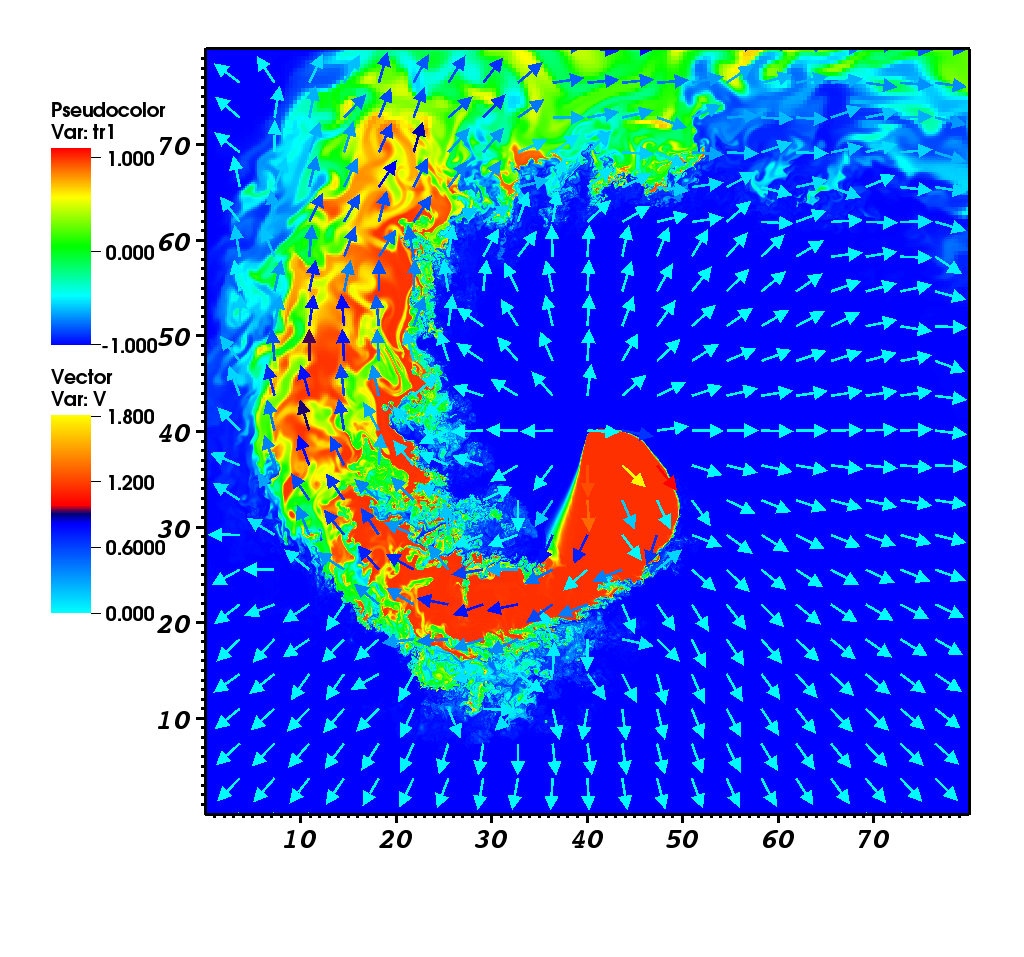}
\vspace{-1cm}
\caption{The same as in Fig.~\ref{rho2} but for the case $\eta=0.6$.}
\label{rho22}
\end{figure}

\begin{figure}
\includegraphics[width=0.5\textwidth,angle=0]{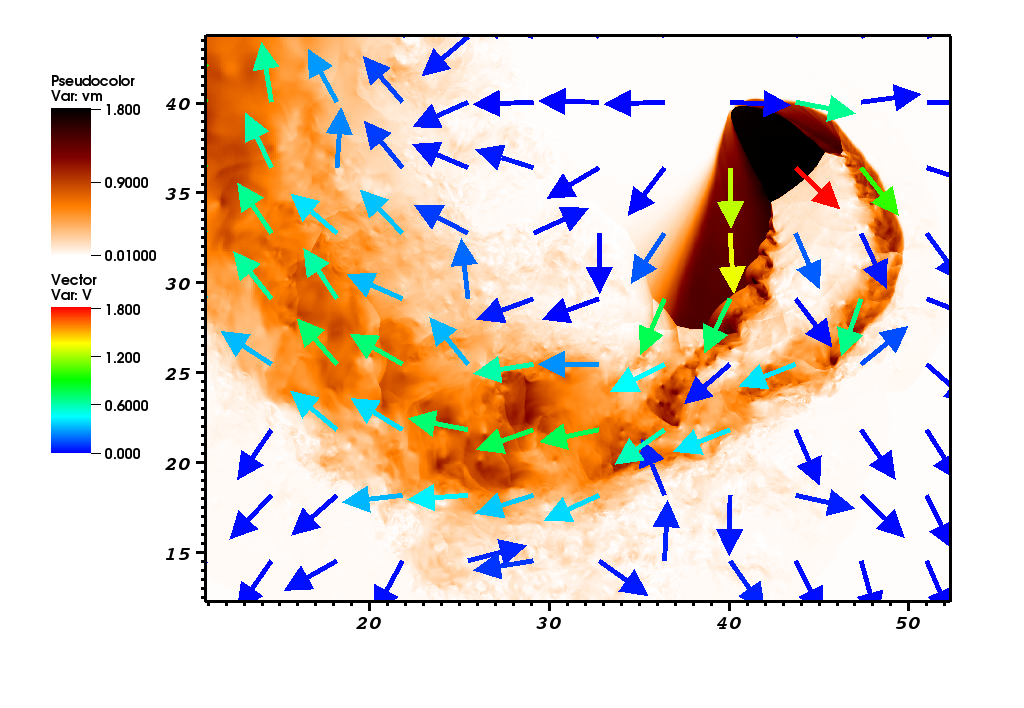}
\vspace{-1cm}
\caption{The same as in Fig.~\ref{rho3} but for the case $\eta=0.6$.}
\label{rho32}
\end{figure}

Figures~\ref{rho1} and \ref{rho12} show the evolution of the density map of the simulated flows during one orbital period,
for $\eta=0.3$ and 0.6, respectively. Physical times earlier than $<0.3\,T$ are not shown since the structure had not yet reached
a quasi-steady configuration on the scales of interest. The snapshots are obtained at $t\simeq0.4\,T$, $0.7\,T$, and
$0.9\,T$. For the last snapshot time, the tracer map is also shown in Figs.~\ref{rho2} ($\eta=0.3$) and \ref{rho22}
($\eta=0.6$), indicating how much pulsar (1) and stellar wind material (-1) are present in the computational cells. Finally,
Figs.~\ref{rho3} ($\eta=0.3$) and \ref{rho32} ($\eta=0.6$) show a map of the module of the four-velocity spatial component
(three-velocity module times the flow Lorentz factor). Arrows and their colours in all figures indicate the direction and module
of the four-velocity spatial component, respectively.

The flow behavior in the inner bow-shaped interaction region is very similar to that found in previous
simulations \citep[e.g.,][]{bog08,bog12}. In particular, the reacceleration of the pulsar wind after the
termination shock \citep{bog08} is also present in our calculations despite the adopted small $\Gamma$-value.
This effect can be noticed in the velocity gradient around the shock apex shown in Figs.~\ref{rho3} and
\ref{rho32}, with bulk Lorentz factors increasing from $\approx 1.1$ to 1.4 (see also Sect.~\ref{w}). On
larger scales, however, the interaction structure starts to depart strongly from being symmetric with respect
to the two-star axis. Several orbital-motion effects are apparent. First, the stellar wind pushes clockwise
the leading side of the interaction structure, forming strong shocks in both the unterminated and the already
terminated pulsar-wind zones.  At the leading edge, and in the unterminated pulsar wind flowing away from the
binary, a strong shock is formed at the distance of $r\sim 5\,a$. On the trailing side, the termination takes
place at around $r\sim 10\,a$. Beyond these points, the shocked pulsar wind is deflected by an angle
$\sim \pi/2$, and becomes strongly turbulent and increasingly loaded with stellar wind material. We note
that, even for highly eccentric systems (such as PSR~B1259$-$63/LS2883), the obtained solution still provides a
qualitative description of the shocked structure. The reason is that the characteristic flow speeds exceed the
pulsar orbital velocity by at least one order of magnitude. This implies, on the one hand, that the flow structure
will adiabatically adjust to the changing $a$ along the orbit. Moreover, the longer orbital timescale
compared to that of the shocked flows will allow the latter to cover a significant fraction of a spiral turn
before $a$ has changed significantly. 

Turbulence is triggered by the Kelvin-Helmholtz (KH) instability in different regions of the two-wind shocked structure. In its
leading edge, the instability grows at the two-wind CD, as also illustrated in Sect.~\ref{w}. The instability couples with
the perturbations generated at the pulsar-wind termination in the opposite direction to the star. At the trailing side, similar
processes take place, although the instability grows substantially further, partially because the flow (re)terminates farther,
partially because the flow lines diverge there. Bending will make the flow lines of the two shocked winds converge again, but it takes
some time for the effect of the Coriolis force to propagate through the shocked structure and cause this convergence.

The tracer maps in Figs.~\ref{rho2} and \ref{rho22} indicate that already within the first orbital turn, i.e. on scales $\sim
(10-100)\,a$, some amount of stellar wind has penetrated deeply into the shocked pulsar flow. This mass-load slows down the
shocked pulsar wind, as shown in Fig.~\ref{rho3}, and in general makes the flow motion more chaotic, and rich both in weak shocks and
turbulence. As seen in the figures, after one orbit the trailing side of the shocked stellar wind already approaches the leading
edge, which has already become turbulent. 

\section{On the pulsar-wind Lorentz factor}\label{w}

As alluded in Sect.~\ref{intro}, approximating the pulsar wind by a non-relativistic flow can underestimate the
pulsar-to-star wind energy-flux ratio, for a given momentum-flux ratio, by a factor of $v/2c$. This can be (roughly) solved
assuming a mildly relativistic pulsar-wind velocity while keeping the Newtonian treatment, which leads to moderately
unphysical results. However, there are additional, more fundamental, differences between relativistic and Newtonian flows
that are relevant to the problem considered here. In this section, we discuss the impact of adopting a relativistic model for
the simulated flows, and the influence of the specific value of the pulsar-wind Lorentz factor. For instance, although the
results obtained in the mildly and highly relativistic cases are similar, the KH instability will tend to develop faster in
the former. The impact of a non-relativistic approximation for the simulated flows is however stronger, since the shocked
flow dynamics will be substantially different when orbital motion is included, even for the mildly relativistic case adopted in
Sect.~\ref{dyn}.

\subsection{The colliding wind region at $\Gamma=10$}\label{g10}

\begin{figure}
\includegraphics[width=0.5\textwidth,angle=0]{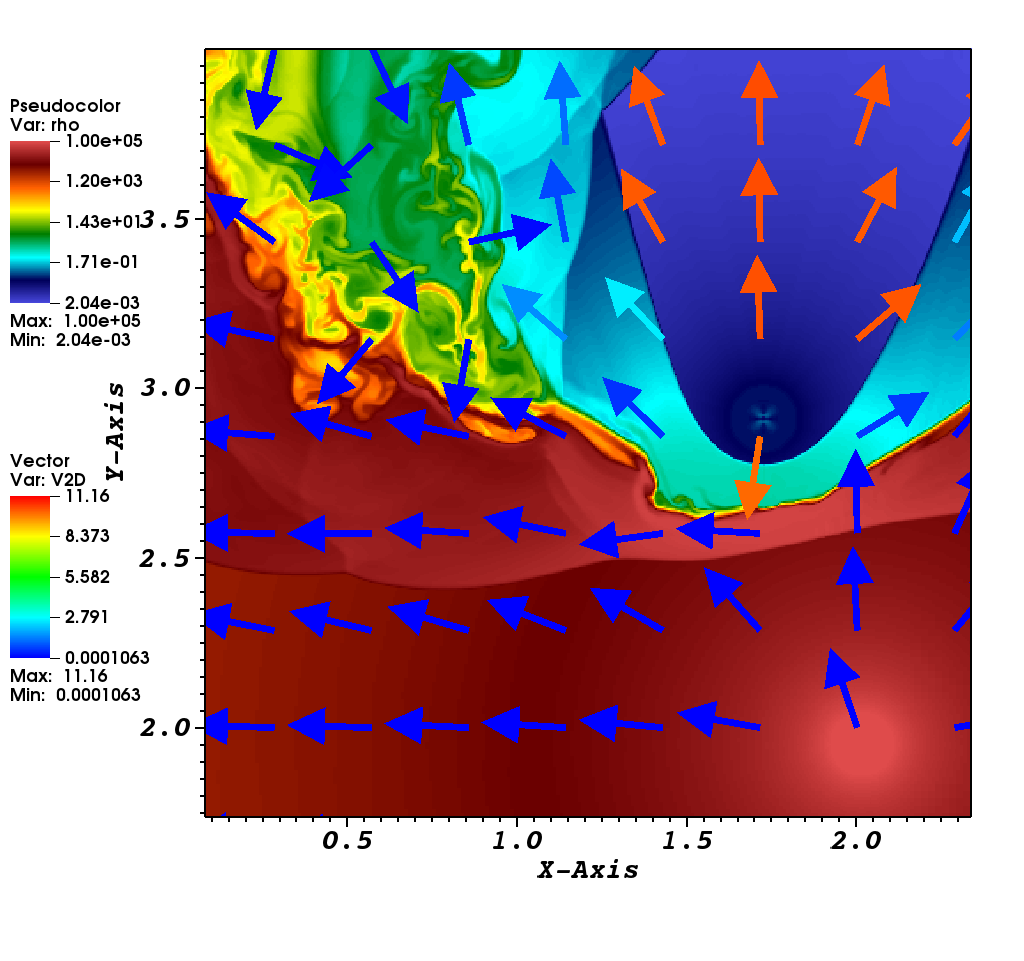}
\includegraphics[width=0.5\textwidth,angle=0]{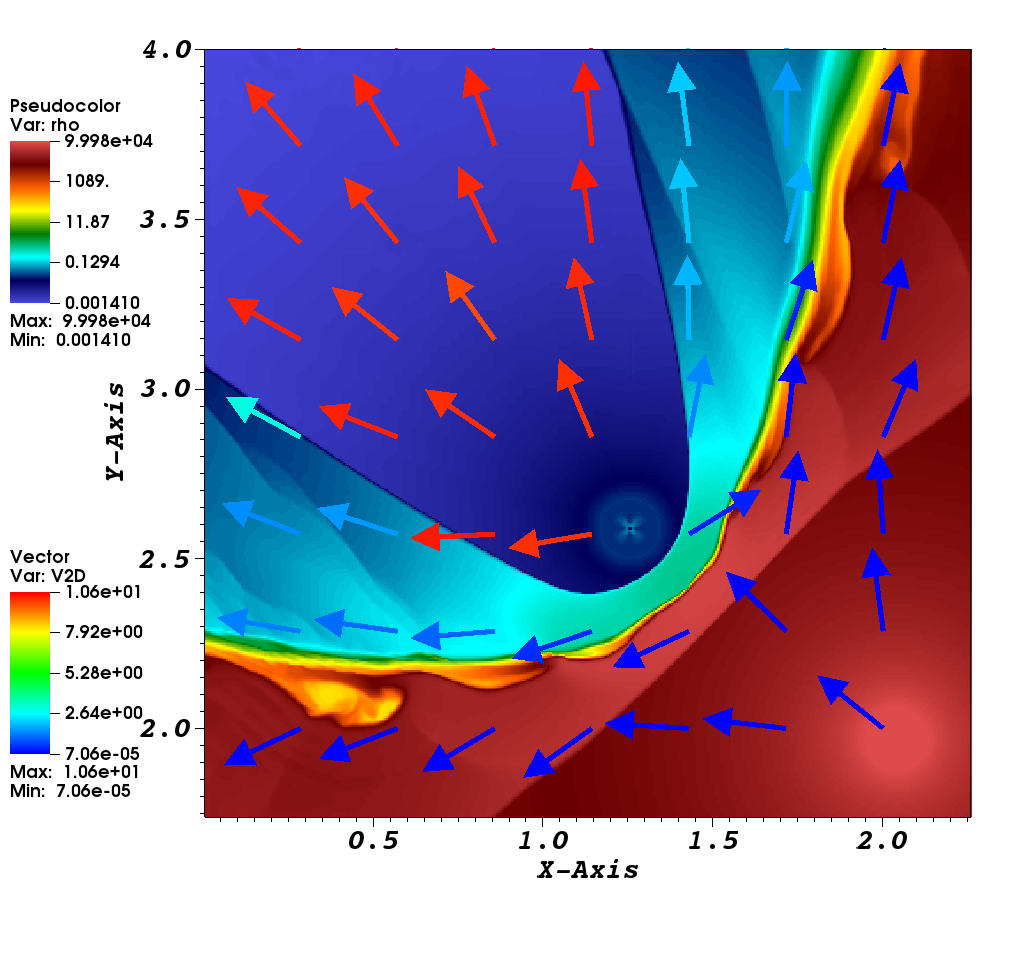}
\vspace{-1cm}
\caption{Top: Density distribution by colour of the interaction region for a 
pulsar wind with $\Gamma=10$ and $\eta=0.3$.
The domain shown is limited to highlight the interaction region.
Bottom: The same as in the top but few hours later, to illustrate that the CD 
can change strongly on timescales much shorter than the binary period.}
\label{g10d}
\end{figure}

\begin{figure}
\includegraphics[width=0.5\textwidth,angle=0]{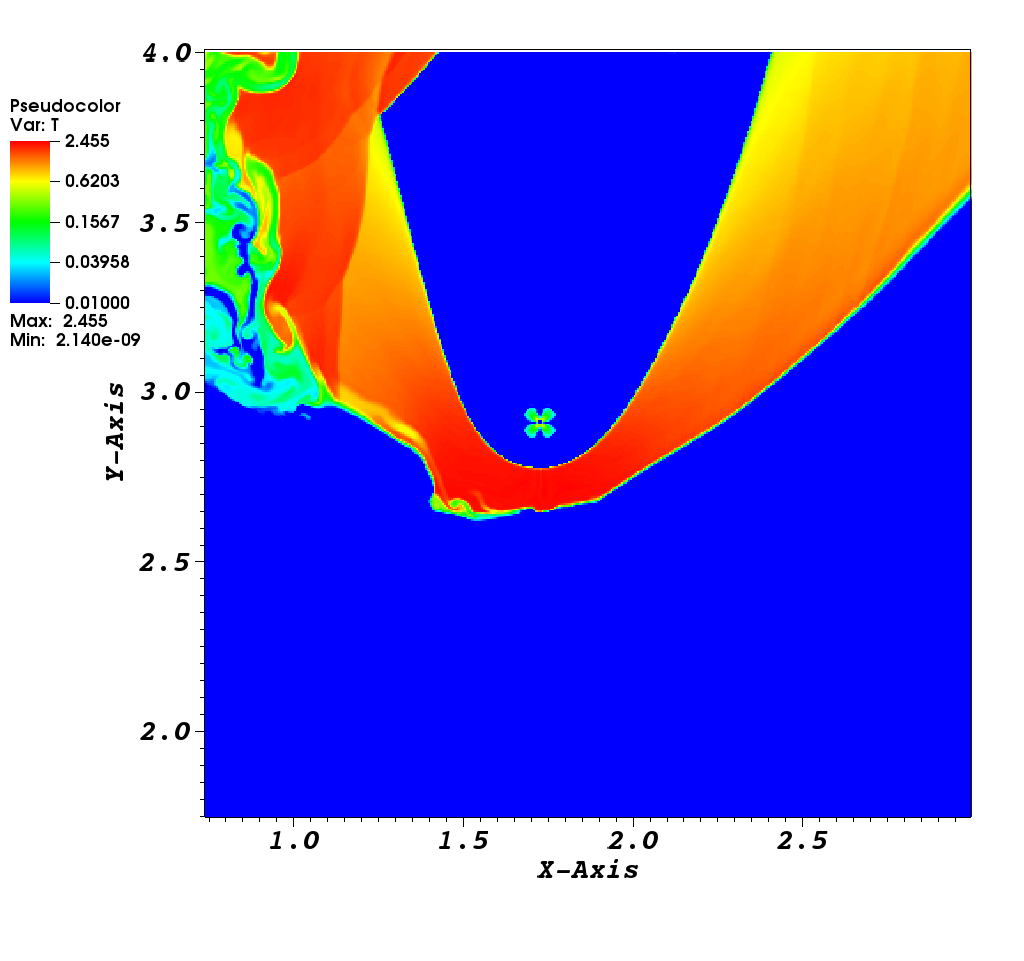}
\vspace{-1cm}
\caption{The same as in Fig.~\ref{g10d} but for the temperature.}
\label{g10t}
\end{figure}

\begin{figure}
\includegraphics[width=0.5\textwidth,angle=0]{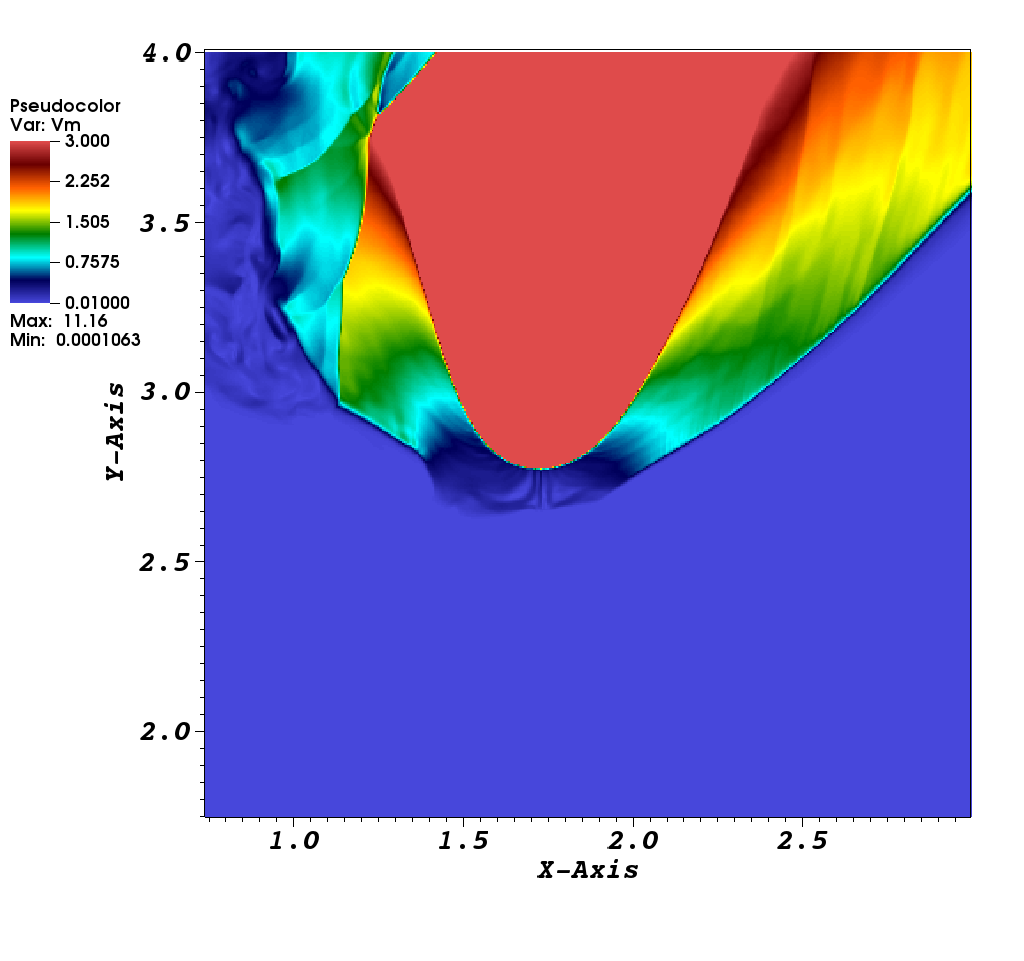}
\vspace{-1cm}
\caption{The same as in Fig.~\ref{g10d} but for the module of the spatial component of the four-velocity.}
\label{g10v}
\end{figure}

We simulated the central part of the interaction region assuming a pulsar wind with $\Gamma=10$ instead of
2. The geometry and system properties were the same as in Sect.~\ref{dyn}, including orbital motion, and
$\eta=0.3$. The simulation was run long enough to reach a quasi-steady state on the considered scales of both $x \in
[0,4\,a]$ and $y \in [0,4\,a]$, with the optical star located at $(2a,2a)$. The adopted value of the wind
Lorentz factor required an effective resolution of the computational grid of about three times the one adopted
in Sect.~\ref{dyn}, now with a domain size of $768\times 768$ cells. The resulting maps for the distribution of
density, temperature, and the spatial component of the four-velocity are shown in Figs.~\ref{g10d}, \ref{g10t}, and
\ref{g10v}. The plot shown in the bottom panel of Fig.~\ref{g10d} illustrates that the CD can change strongly
on timescales much shorter than the binary period. The results of this numerical test show that strong shocks
induced by the orbital motion are still present when the pulsar-wind Lorentz factor is higher, and the
reacceleration of the shocked pulsar wind is stronger in the case of $\Gamma=10$ than for $\Gamma=2$. As shown
in Sect. \ref{rni}, the shocks induced by the orbital motion are a general feature when adopting
a relativistic flow model, and they should also be present in the ultra-relativistic case. Regarding the
development of the KH instability at the CD, in the considered case the growth rate does not seem to be significantly
reduced. However, this conclusion may not be generalized in a straightforward way to the ultra-relativistic
case, which is considered in the next section.

\subsection{Instability growth at the CD and $\Gamma$}\label{slab}

Although the shocked structure evolution is the result of the interplay of different mechanisms operating
on different scales, the KH instability developing in the CD can by itself have a strong impact on the flow
structure. To study the instability growth in the CD, we performed 2D test simulations of the shocked pulsar and
stellar winds adopting a mildly and an ultra-relativistic pulsar wind. The context of the simulations is
the scenario described in Sect.~\ref{dyn}. The calculations were performed with a relativistic hydrodynamical
code whose characteristics are described for instance at the end of Sect.~2 in \cite{per05}. The size of the
computational box was $10^{12}{\,\rm cm}\times 3.3\cdot10^{11}{\,\rm cm}$, and the resolution, $1920\times 640$
cells. The two shocked winds are homogeneous in the direction perpendicular to the simulated plane, and are
initially in thermal pressure balance. It was assumed that the shocked pulsar wind had already been
reaccelerated, thus the thermal pressure was significantly lower than the ram pressure. The shocked
stellar wind was located at the upper half of the computational box (see below) and was assumed to be initially
at rest. 

\begin{figure}
\includegraphics[width=0.5\textwidth,angle=0]{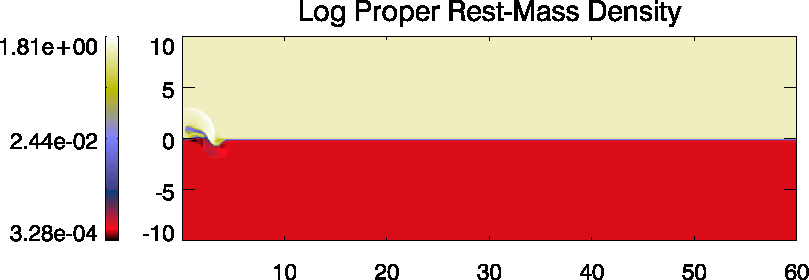}
\includegraphics[width=0.5\textwidth,angle=0]{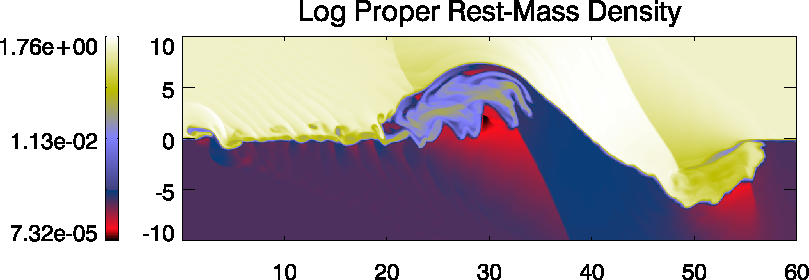}
\caption{Density distribution by colour in arbitrary units of the development of instabilities at the 
contact discontinuity after 2.8 (top) and 43~s (bottom) from the beginning of the 
simulation, for a mildly relativistic pulsar wind. The axes units
are $1.7\times 10^{10}$~cm.}
\label{slab1}
\end{figure}

\begin{figure}
\includegraphics[width=0.5\textwidth,angle=0]{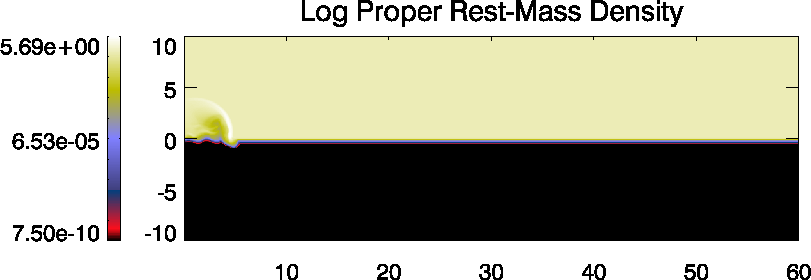}
\includegraphics[width=0.5\textwidth,angle=0]{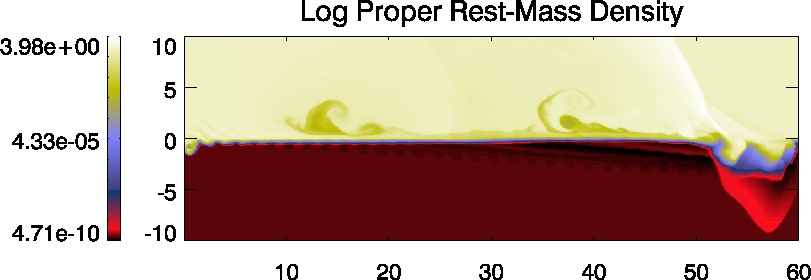}
\caption{The same as in Fig.~\ref{slab1} but after 2.8 (top) and 34~s (bottom), 
for the case with a bulk Lorentz factor of the shocked pulsar wind 
of $10$ (in the laboratory frame), and a microscopic Lorentz factor for the particles of $10^5$ 
(in the flow frame).}
\label{slab2}
\end{figure}

The results for the density distribution illustrating the growth of seed perturbations are presented in
Figs.~\ref{slab1} and \ref{slab2}. The pulsar wind in Fig.~\ref{slab1} has a bulk Lorentz factor of 2, and a
thermal pressure of about 10\% of the ram pressure. In Figure~\ref{slab2}, the pulsar-wind bulk Lorentz factor
is 10, and the microscopic Lorentz factor in the flow frame, related to temperature, is $10^5$. This case is
very different from the case considered in Sect.~\ref{g10}, and would roughly correspond to a shocked and
reaccelerated ultra-relativistic wind with (unshocked) $\Gamma\sim 10^6$.  A comparison of Figs.~\ref{slab1}
and \ref{slab2} shows that, in the ultra-relativistic case, the development of the KH instability is
significantly weaker, which is  consistent with earlier numerical studies. In particular, a discussion of the
reduction in the instability growth and related non-linear effects (namely mixing, disruption, and
deceleration) for high Lorentz-factor flows can be found in \cite{per04a}, \cite{per04b}, and \cite{per05}, in
the context of planar and axisymmetric cylindrical jets.

Although the situation considered in this study has been very much simplified, one may still draw the
conclusion that the shocked structure in high-mass binaries hosting a pulsar is less likely to develop the KH
instability than massive star binaries. However, even in the case of an ultra-relativistic flow, the figures show 
that there is still a
significant growth of the KH instability on scales comparable to those of the binary system. This, plus
the strong asymmetry of the interaction introduced by orbital motion, makes the generation of shocks and
turbulent flows at the CD very likely on scales of $\sim a$, regardless of the value of $\Gamma$.

\subsection{Relativistic vs Newtonian inertia}\label{rni}

Orbital motion introduces a strong non-stationary component in the dynamics of the shocked structure. In
such a case, the inertia of the flow has a very strong influence, and there is a major difference between the
relativistic and Newtonian case: the relativistic inertia is proportional to pressure when this is not
negligible compared to the rest mass energy (as expected here), whereas Newtonian inertia is proportional to
density. Therefore, although Newtonian simulations may reproduce the geometry of the interaction region within
the system, where the orbital motion is negligible, on scales $\gtrsim a$ the Coriolis force starts to deflect
the flow, and inertia then becomes important. 

If one of the shocked winds is at least mildly relativistic, for it to be significantly deflected the exerted 
lateral pressure is to be a
substantial fraction of the ram pressure of the flow. Since the plasma flowing out from the system is already trans-sonic,
this lateral pressure will be ($\Gamma^2\times$) 
higher than the thermal pressure, and the flow deflection will produce shocks. These shocks, which are
stronger for higher $\Gamma$, will trigger large CD perturbations, and can be a source of energy dissipation in the form of
particle acceleration. 

In a Newtonian flow with the momentum and energy fluxes of the shocked pulsar wind, density and therefore inertia
would be very small even if the plasma were relatively hot. Such a fluid could be smoothly deflected
by sound waves, rendering a much more stable structure without much non-thermal activity. This conclusion
also applies if the wind has a non-relativistic velocity but the same momentum flux, and thus smaller energy
flux, than its relativistic counterpart. 

The effects discussed here balance the higher stability of the CD with an ultra-relativistic pulsar wind
discussed in Sect.~\ref{slab}. By itself, the instability growth rate would depend on (small) seed
perturbations of the CD of some sort. However, under orbital motion, strong non-linear perturbations are
unavoidable, as shown in Sect.~\ref{g10}. We note that this conclusion is based on the relativistic nature of the
flow, and does not apply to massive star binaries with colliding winds.

\section{Discussion}\label{disc}

The simulation results show that the colliding pulsar and stellar winds produce a coherent flow with low
entropy up to only a few $a$.  Beyond that point, for stellar and pulsar winds typical for gamma-ray binaries
(possibly) hosting a non-accreting pulsar, the shocked flow becomes bent, non-ballistic, and highly turbulent,
with strong wind mixing. It seems therefore unlikely that the shocked structure will remain as a spiral on
scales $\gg a$ (even for $\eta=0.6\lesssim 1$). As seen by comparing the cases with $\eta=0.3$ and 0.6, the
smaller the two-wind momentum flux ratio, the earlier the shocked structure is closed and disrupted by the
orbital motion. Overall, the flow evolution is quite similar to earlier analytical predictions \citep{bb11}.

The adopted $\Gamma$-value already captures the main physics of the colliding wind region and the shocked
structure formed there. Higher wind Lorentz factors may yield more stable flows at the CD in an idealized
situation, but the impact of orbital motion enhances instability and triggers shocks, and overall the global
picture should not change dramatically between $\Gamma=2$ and $\Gamma\gg 1$. Newtonian simulations
thus seem suitable only for the colliding wind region on its smallest scales, within the binary, because further out
the impact of orbital motion will be much smoother than in the relativistic case, missing much of the instability
and shock generation. In addition, a non-relativistic pulsar wind, unlike its relativistic counterpart, could
not re-accelerate and its speed could not then again approach $c$ after being shocked.

In addition to the pulsar-wind shock facing the star, and the unshocked pulsar wind itself, the present calculations reveal
additional sites of non-thermal activity \citep[as hinted already in][]{bb11}. Most of the pulsar-wind kinetic energy that
is not reprocessed on the binary system scales is converted into internal energy farther out, either through strong shocks in
the direction away from the star, or through weaker shocks and turbulence further downstream. Given their proximity to the
binary system, the strong shocks {\it behind} the pulsar seem ideal sites for particle acceleration and high-energy emission.
For instance, farther from the binary, the expected weaker synchrotron and inverse Compton losses would allow the accelerated
particles to reach higher energies. This could explain the very energetic photons seen from LS~5039 \citep{kha08}. In
addition, an emitter farther from the star would allow gamma rays to avoid severe absorption in the stellar photon field
\citep[e.g.,][]{bos08}. Finally, the presence of emitting regions with different physical conditions may explain the
broad-band spectrum in some gamma-ray binaries (e.g. \citealt{tak09}; Zabalza et al., in preparation).

The complex hydrodynamics of the shocked flow on scales $\gg a$, with weak shocks and turbulence, indicates
that radio-emitting electrons may be accelerated there instead of at the termination shock facing the star,
where free-free radio absorption can be severe, and the high energies of the accelerated electrons
\citep[e.g.,][]{ken84} are unsuitable for efficient radio emission. The radio electrons will initially move
at mildly relativistic speeds, at least in most of the emitting flow, but for $r\gtrsim 100\,a$ further mixing
will likely slow down the flow. The spatial scale of the simulated flow corresponds to $\sim
10-100$~milliarcseconds at $\sim 2$~kpc distance, so the production and evolution of the radio-emitting
electrons, and the structure of the underlying flow, can be studied with current VLBI instrumentation
\citep[see][]{dha06,rib08,mol11,mol11b}. 

The mildly relativistic speeds downstream of the strong shocks in the pulsar wind imply that Doppler boosting can
be significant there (as proposed in \citealt{kha08b}; see also, e.g., \citealt{dub10}). However, we must
emphasize that the motion of the whole interaction structure is non-ballistic, with rather complex velocity
distributions both in terms of module and direction. This should induce through Doppler boosting complicated variability
patterns with different timescales in the non-thermal emission. This variability would appear on top of that
intrinsic to the non-thermal activity itself.

\begin{acknowledgements} 
We thank an anonymous referee for useful and constructive suggestions.
The calculations were carried out in the cluster of Moscow State University {\it Chebyshev}. We thank Andrea Mignone and the
{\it PLUTO} team for the possibility to use the {\it PLUTO} code. We also thank the {\it Chombo} team for the possibility to
use the {\it Chombo} code. 
The visualization of the results was done using the VisIt package.
The research leading to these results has received
funding from the European Union Seventh Framework Program (FP7/2007-2013) under grant agreement PIEF-GA-2009-252463. V.B.-R.
acknowledges support by the Spanish  {\it Ministerio de Ciencia e Innovaci\'on} (MICINN) under grants AYA2010-21782-C03-01
and FPA2010-22056-C06-02, and MP acknowledges 
support by the Spanish {\it Ministerio de Ciencia e Innovaci\'on} (MICINN) grants
AYA2010-21322-C03-01, AYA2010-21097-C03-01 and CONSOLIDER2007-00050.  
\end{acknowledgements}

\bibliographystyle{aa}
\bibliography{text}
\end{document}